\documentclass[aps,prl,reprint,superscriptaddress]{revtex4-1}

\usepackage{graphicx}
\usepackage{amssymb,amsmath}
\usepackage{epstopdf}
\usepackage{float}
\usepackage{xcolor}
\usepackage{pifont}
\usepackage{bm}
\usepackage{mathrsfs}
\usepackage{bbold}

\usepackage{times}
\usepackage{url}

\begin{document}

\title{Characterization of the magnetic field through the three-body loss near a narrow Feshbach resonance}

\author{Yang Chen}
\thanks{These authors contributed equally to this work.}
\affiliation{School of Physics and Astronomy, Sun Yat-sen University, Zhuhai, Guangdong, 519082, China}

\author{Shuai Peng}
\thanks{These authors contributed equally to this work.}
\affiliation{School of Physics and Astronomy, Sun Yat-sen University, Zhuhai, Guangdong, 519082, China}

\author{Hongwei Gong}
\affiliation{School of Physics and Astronomy, Sun Yat-sen University, Zhuhai, Guangdong, 519082, China}
\author{Xiao Zhang}
\affiliation{School of Physics and Astronomy, Sun Yat-sen University, Zhuhai, Guangdong, 519082, China}

\author{Jiaming Li}
\email[]{lijiam29@mail.sysu.edu.cn}
\affiliation{School of Physics and Astronomy, Sun Yat-sen University, Zhuhai, Guangdong, 519082, China}
\affiliation{Center of Quantum Information Technology, Shenzhen Research Institute of Sun Yat-sen University, Nanshan Shenzhen, Guangdong, China 518087}
\affiliation{Guangdong Provincial Key Laboratory of Quantum Metrology and Sensing, Zhuhai 519082, China}

\author{Le Luo}
\email[]{luole5@mail.sysu.edu.cn} 
\affiliation{School of Physics and Astronomy, Sun Yat-sen University, Zhuhai, Guangdong, 519082, China}
\affiliation{Center of Quantum Information Technology, Shenzhen Research Institute of Sun Yat-sen University, Nanshan Shenzhen, Guangdong, China 518087}
\affiliation{Guangdong Provincial Key Laboratory of Quantum Metrology and Sensing, Zhuhai 519082, China}
\date{\today}

\begin{abstract}
The narrow s-wave Feshbach resonance of a $^6$Li Fermi gas shows
strong three-body loss, which is proposed to be used to measure the
minute change of a magnetic field around the resonance. However, the
eddy current will cause ultracold atom experiencing a magnetic field
delayed to the desired magnetic field from the current of the
magnetic coils. The elimination of the eddy current effect will play
a key role in any experiments that motivated to measure the magnetic
field to the precision of a part per million stability. Here, we
apply a method to correct the eddy current effect for precision
measurement of the magnetic field. We first record the three-body
loss influenced by the effect of induced eddy current, then use a
certain model to obtain the time constant of the actual magnetic
field by fitting the atom loss. This precisely determines the actual
magnetic field according to the time response of the three-body
loss. After that, we implement the desired magnetic field to the
atoms so that we can analyze the three-body loss across the whole
narrow Feshbach resonance. The results show that the three-body
recombination is the dominated loss mechanism near the resonance.
We expect this practical method of correcting the eddy current error
of the magnetic field can be further applied to the future studies
of quantum few- and many-body physics near a narrow Feshbach
resonance.
\end{abstract}

\maketitle


A narrow Feshbach resonance is a closed channel dominated resonance rather than a open channel dominated resonance as in the broad case, whose near-threshold scattering and bound states only exist over a small fraction of width near resonance~\cite{Chin2010,Blackley2014}.
It usually has a nontrivial energy dependent collisional phase shift as well as a narrow resonance width,
which makes the theoretical modeling and experimental observation become complex and difficulty~\cite{Wang2011, Ho2012,Gao2010,Petrov200401,Suno2003,Hazlett2012,Wang2013,Yoshida2018,Gaebler2007,Arunkumar2018,Kohstall2012,Green2020}.
But this also introduces many interesting and difference physics in compare to the widely studied broad Feshbach resonance. For example, the atomic-dimer relaxation ratio of the broad Feshbach resonance in an ultracold $^6$Li Fermi gas is suppressed as $a^{3.33}$ for two-body s-wave scattering length $a>0$~\cite{Petrov2004}, which results in an stable atom and molecular mixture, while this inelastic process is predicted to be enhanced if the effective range of the s-wave scattering phase shift $r_{\mathrm{eff}}$ is larger than $a$~\cite{Wang2011}.

The two lowest-energy hyperfine ground states mixture of $^6$Li ($\left| \uparrow \right> $ and $\left| \downarrow \right> $) has a narrow s-wave Feshbach resonance at 543.3 G, whose resonance width is estimated to be only 0.1 G and  $r_{\mathrm{eff}}=-71000 a_0$ of the interaction potential is larger than the interparticle separation\cite{Hazlett2012}. Thus, the three-body recombination is supposed to be strong. As our previous study shows that three-body recombination of  $^6$Li through the narrow Feshbach resonance follows a van der Waals universal and has an asymmetric strength profile across the resonance~\cite{Li2018}.
In the Bardeen-Cooper-Schrieffer (BCS) side, the loss width is about $E_F/\mu_B$, where $E_F$ is the Fermi energy and $\mu_B$ is the Bohr magneton.
Therefore, in an extreme low temperature, the three-body loss will be located in a very small magnetic field regime with strong strength.
In a sense, a milligauss stability and millisecond speed of magnetic field are suggested.

In this letter, we propose to use the ultra narrow width feature of the three-body loss near the $^6$Li narrow Feshbach resonance to sense the magnetic field. This method is a straight forward way to characterize the experienced magnetic field of the ultracold atoms. By applying the sensitive magnetic field discriminator from the three-body loss in narrow Feshbach resonance, the magnetic field changing time constant is highly determined in our system. Then we use the fitted result to study the time evolution result of the atom loss in the Bose-Einstein condensate (BEC) side and find that the main collision mechanism is also the three-body
recombination in the BEC side when the magnetic field is close to resonance. This is also the first experimental measurement.

\begin{figure}[htbp]
    \begin{center}
        \includegraphics[width=\columnwidth, angle=0]{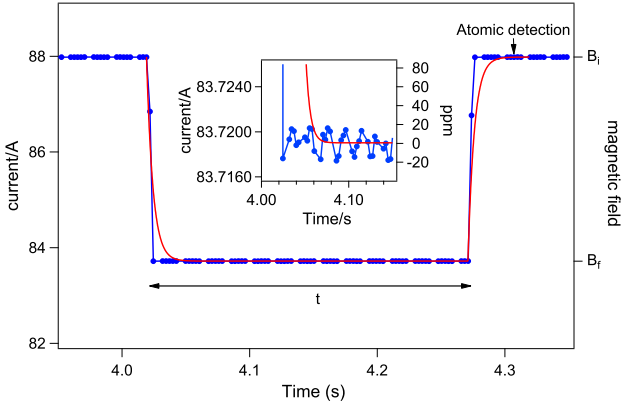}
        \caption{Timing sequence of driving current in the coils (blue dots) and estimated bias magnetic field at the atom place (red curve). The driving current is measured by a 7.5 bit digital multimeter (Keithley DMM7510) with 0.1 number of power line cycles and auto-zero setup. The inner figure zooms in the current dynamic at the turning point. The real magnetic field is calculate from Eq.~(\ref{eq:B-t}) with the measured $\tau=5$ $\mathrm{ms}$, $B_i=570.9913$ $\mathrm{G}$, and $B_f=543.2735$ $\mathrm{ G}$. In a typical three-body loss measurement sequence, the magnetic field starts from $B_i$, then jumps to $B_f$ and stay their for a variable time $t$, after that jumps back to $B_i$ to do atomic detection.
        }    \label{p:B_Field_Timing}
    \end{center}
\end{figure}

We product the ultracold Fermi gases in an optical dipole trap~\cite{Li2018,Li2016}. After the gases is cooled to $0.15$ $\mathrm{\mu K}$, about $N(t=0) = 32500$ atoms per spin is left.
The experimental timing sequence of magnetic field is shown in Fig.~\ref{p:B_Field_Timing}. We
sweep the magnetic field to a initial value $B_i$ = 570 G,
where is tested to be stable for two component Fermi
gases~\cite{Hazlett2012,Li2018}.
Then the magnetic field is jumped to the target value $B_f$, stay for a variable time $t$.
After that it is jumped back to $B_i$ and do the atom number $N(t)$ detection.
As a consequence of the induced eddy current in the metal of our cold atom apparatus, the time response of magnetic field at the atom place is slower than the response of the driving current in the coils, which introduce error for many magnetic field dependent measurements if we use the driving current to scale the actual magnetic field.
Especially, in our narrow Feshbach resonance experiments, the stabilization of the magnetic field often required to be a part per million (ppm) level, which put forward higher requirements for the dynamic properties of the magnetic field.
Fig.~\ref{p:B_Field_Timing} presents the measured driven current of the coils and estimated magnetic field during a fast sweep.
The result shows the driving current can reach 20 ppm in 5.4 ms, but a better current resolution is absent due to the measurement limited of the digital multimeter.
Considering the induced eddy current, the slowed down magnetic field response can be expressed by a first-order step response model~\cite{Michele2008}, which is expressed as
\begin{equation}\label{eq:B-t}
B(t)=(B_i-B_f)e^{-\frac{t}{\tau}}+B_f
\end{equation}
Where $\tau$ is the time constant. The expected value of $\tau$ is zero in an ideal scenario, but in our system, its value is on the order of milliseconds.

\begin{figure}[htbp]
    \begin{center}
        \includegraphics[width=\columnwidth, angle=0]{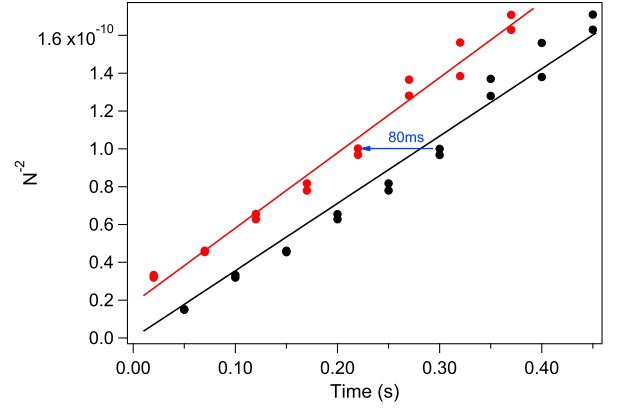}
        \caption{Typical time evolution of $1/N^2$ with a three-body loss in our experiment. Red dots are the 80 ms time shifted data of black dots. If we use a linear function to fit the black dots, the fitted $1/N^2(t=0)$ will become almost zero, which is because of the very beginning data are taken under an unstable magnetic field due to the induced eddy current. Instead, fitting the red dots will give the right $L_3$ and $1/N^2(t=0)$.}    \label{p:L_3_measurement}
    \end{center}
\end{figure}

\begin{figure*}[htbp]
    \begin{center}
        \includegraphics[width=2\columnwidth]{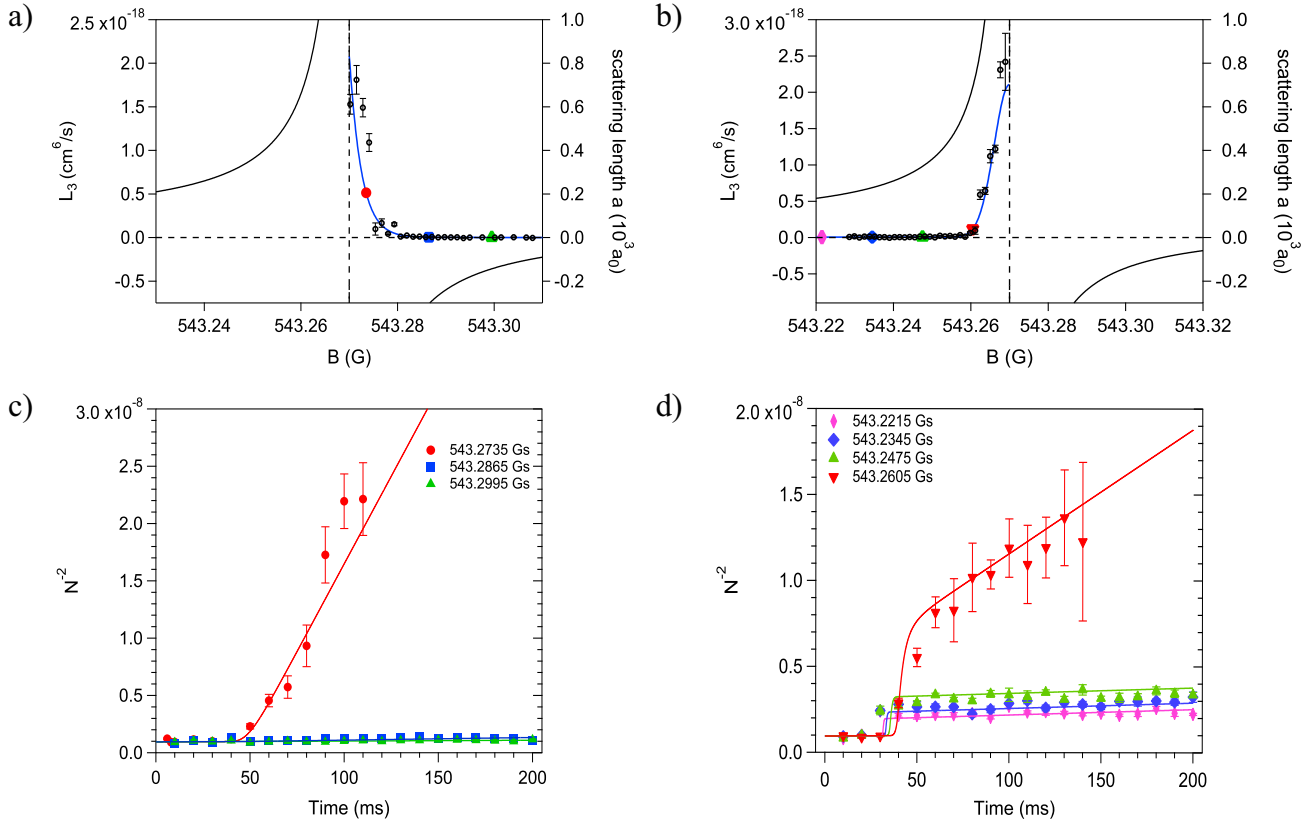}
        \caption{(a) measured $L_3(B)$ at the BCS side. Blue curve is the fitted result with Eq.~(\ref{eq:L3_B}).  (c) time-dependent $1/N^2$ at three different $B_f$ values, which are marked at (a). (b) measured $L_3(B)$ at the BEC side. Blue curve is the fitted result with Gaussian profile. (d) time-dependent $1/N^2$ at four different $B_f$ values, which are marked at (b). Both the width of $L_3(B)$ are much smaller than the resonance width (Black curve is the calculated scattering length). Solid lines in (b, d) are the fitting curves with Eq.~\ref{eq:N-L3_solve} and the fitted time constant of the magnetic field curve.
        }  \label{p:fitting_result}
    \end{center}
\end{figure*}

Three-body recombination at the BCS regime is measured. Its magnetic dependent properties can be described by\cite{Li2018}
\begin{equation}\label{eq:L3_B}
L_3(B)=\frac{3h^3K_{ad}}{(\pi m k_B T)^{3/2}}
\text{exp}[-\frac{2\mu_B(B-B_0)}{k_B T}]
\end{equation}
where $K_{ad}$ is the atom-dimer relaxation rate, $B_0$ is the resonant magnetic field, $h$ is the Planck constant and $k_B$ is the Boltzmann constant.
Accounting the slow magnetic response, the three-body atom loss $N(t)$ is modified to
\begin{equation}\label{eq:N-L3_solve}
\frac{1}{N^2(t)}=\frac{1}{V^2}\,
\int_0^t L_3(B(t))dt+\frac{1}{N^2(t=0)}
\end{equation}
Where $V$ is the average volume of atom gas.
It is very important that we find $L_3$ is not sensitive to the very beginning unstable magnetic field response.
So we submit the time of each measurements with a large enough time span, like 80 ms in our experiment, then do the fitting and get $L_3$.
Fig.~\ref{p:L_3_measurement} presents these processes.
We need point out that although we derive $L_3$, we still loss some information at the very beginning, which prevent our to explore some rapid phenomena, such as molecular formations\cite{Strecker2003}, atomic-molecular collision and molecular-molecular collision~\cite{Wang2013}, etc.
By fitting the time shifted $1/N^2(t)$ data, we can get the $L_3$. Fig. ~\ref{p:fitting_result}(a) shows the measured $L_3(B)$ at the BCS regime, which follows the tendency described in Eq.~\ref{eq:L3_B} with a fitted $K_{ad}=4.00\times 10^{-8}$ $\mathrm{cm}^3/\mathrm{s}$. 
Note that the $1/e$ width of $L_3(B)$ is only $0.0023$ G, which is a much smaller than the resonance width 0.1 G.
We choose three different $B_f$ values, and measured their $N(t)$ respectively, as shown in Fig.~\ref{p:fitting_result}(c).
We manually fit $N(t)$ with Eq.~(\ref{eq:N-L3_solve}), get the time constant of actual magnetic field response $\tau=5$ $\mathrm{ms}$.
In this way, the actual magnetic field need about 54 ms to reach 1 ppm range, which also indicates this delay effect should be considered in many rapid experiments.
We further use this magnetic field response prediction the $N(t)$ at the BEC side, as shown in Fig.~\ref{p:fitting_result}(b, d). The results turn out they are also in good agreement with the theoretical calculation.

It is notice that the step changing of $1/N^2$ around 35 ms in Fig.~\ref{p:fitting_result}(d) is the integration of three-body loss from the BCS side to part of the BEC side, rather than the evidence of the molecular formation. And the different final values of $1/N^2$ in the deep BEC regime are because of the different
overlapping strength between the $L_3(B)$ and the exponential decay of magnetic field under different $B_f$. In a sense, the precision determination of the actual magnetic field is very important.
Furthermore, our studies show the three-body loss is the domination lossy
mechanics in the BEC side for a narrow Feshbach resonance, which is also consistent with the theory prediction~\cite{Wang2011}.
It is should be pointed out that although our results shows the main feature across the narrow s-wave Feshbach resonance is three-body recombination, the profile of $L_3(B)$ in the BEC regime is unknown and without experimental measurement before. Here, we use a Gaussian profile to fit the $L_3(B)$ as shown in Fig.~\ref{p:fitting_result}(b). The reason is experiment data of $L_3(B)$ has a very narrow unitary regimes at the resonance point, which may have some connections with the narrow p-wave Feshbach resonance~\cite{Waseem2018}. The fitted Gaussian width is about 3 mG, which is comparable with the width at the BCS side.


In summary, we use the narrow Feshbach resonance of $^6$Li Fermi
gases to realize a precision measurement of the magnetic field near
the resonance. Our method directly measures the magnetic field at
the location of the atom cloud, eliminating the effects of induced
eddy current and residual magnetic. According to the result, the
magnetic field response shows a large delay in comparing with the
driven current. This is a key factor to be considered in many
experiments related to the narrow Feshbach resonance. We find that
three-body recombination dominates the atom loss in both BEC and BCS
regime of the resonance. In the degenerate temperature, the energy
broadening of the resonance is small. The technique of eliminating
the eddy current provides a way to precisely determine the magnetic
field of the narrow Feshbach resonance, and will give other chances
for the future studies.

\section*{Acknowledgements}
This work is supported by Key-Area Research and Development Program
of GuangDong Province under Grant No. 2019B030330001. J. Li received
supports from National Natural Science Foundation of China (NSFC)
under Grant No. 11804406, Fundamental Research Funds for Sun Yat-sen
University 18lgpy78, Science and Technology Program of Guangzhou
2019-030105-3001-0035. L. Luo received supports from NSFC under Grant
No. 11774436, Guangdong Province Youth Talent Program under Grant
No. 2017GC010656, Sun Yat-sen University Core Technology Development
Fund.

%

\end{document}